\documentstyle[12pt]{article}

\renewcommand{\thefootnote}{\fnsymbol{footnote}}
\newcommand{\EQ}{\begin{equation}}
\newcommand{\EN}{\end{equation}}
\newcommand{\bea}{\begin{eqnarray}}
\newcommand{\ena}{\end{eqnarray}}
\newcommand{\vs}[1]{\vspace{#1 mm}}

\newcommand{\uda}{\nearrow \kern-1em \searrow}

\newcommand{\CMP}[1]{Comm.\ Math.\ Phys.\ {\bf #1}}

\def\jm#1#2{\vert #1,#2 >}
\def\qint#1{[ #1 ]}
\def\qinth#1{[ #1 ]^{1/2}}
\def\threej#1#2#3#4#5#6{\left[ \matrix{ {#1}&{#2}&{#3} \cr
{#4}&{#5}&{#6}
\cr}\right]_q}
\def\ket#1{\vert #1>}
\def\oof{{ 1 \over 4} }
\def\oot{{ 1 \over 2} }
\def\jjm#1#2#3{ \vert #1,#2,#3 >}
\def\jph{j+1/2}
\def\jmh{j-1/2}
\def\ep{{e_1}}
\def\ez{{e_0}}
\def\emi{{e_{-1}}}
\def\bp{{b_{1/2}}}
\def\bm{{b_{-1/2}}}
\def\cp{c_{1/2}}
\def\cm{c_{-1/2}}
\def\qh{{q^{1/2}}}
\def\qmh{{q^{-1/2}}}
\def\qq{{q^{1/4}}}
\def\qmq{{q^{-1/4}}}
\def\qhmh{{(\qh - \qmh)}}

\begin{document}

\topmargin 0pt
\oddsidemargin 5mm

\begin{titlepage}
\setcounter{page}{0}
\begin{flushright}
OU-HET 191 \\
% hep-th/9405180 \\
May, 1994
\end{flushright}

\vs{15}
\begin{center}
{\Large DIRAC OPERATORS ON QUANTUM TWO SPHERES}
\vs{15}

{\large Kazutoshi Ohta\footnote{e-mail address: kohta@oskth.kek.jp} and Hisao
Suzuki\footnote{e-mail address: suzuki@fuji2.wani.osaka-u.ac.jp}}\\
\vs{8}
{\em Department of Physics, \\
Osaka University \\ Toyonaka, Osaka 560, Japan} \\
\end{center}
\vs{10}

\centerline{{\bf{Abstract}}}
We investigate the spin $1/2$ fermions on quantum two spheres.
It is shown that the wave functions of fermions and a Dirac Operator on
quantum two spheres can be constructed in a manifestly covariant
way under the quantum group $SU(2)_q$.
 The concept of total angular momentum and chirality can be
expressed
by using $q$-analog of Pauli-matrices and appropriate commutation
relations.

\end{titlepage}
\newpage
\renewcommand{\thefootnote}{\arabic{footnote}}
\setcounter{footnote}{0}

The structure of quantum groups \cite{Dri}\cite{Jim}\cite{Wor} often appears as
hidden
symmetry in physics. One of interests in the quantum groups is that we
can
describe quantum manifolds as a non-commutative geometry.
\cite{Man}\cite{Pod}\cite{WZ}\cite{CSSW}
The quantum manifolds have been studied as possible deformation of
manifolds
along with the geometrical understanding of the quantum groups. And
that
deformation can be used to generalize the ordinary notion of
space(time)
symmetry.

Recently, it was suggested that compact quantum manifolds can be used
as
an regularization of field theories.  It was shown\cite{Suzuki} for the scalar
field on the quantum $2,3$ and $4$
spheres\cite{Pod}\cite{RTF} that   the number of wave
functions can be finite when $q$ is a root of unity $(q^l =1).$ The
parameter $l$
is determined by the ratio of the scale of the universe and that of the
ultraviolet cut-off.  Another attractive feature of the quantum
spheres is the scale dependence. The effects of the deformed manifold
are
invisible at sufficient low energy scale.  In other words, the quantum
manifolds might (effectively) describe the strong quantum fluctuations
of  the manifolds in
microscopic level, which  can be smoothed out in our experimental
scale.

Although the scalar fields on quantum spheres are constructed
consistently, the consistency of the higher spin fields has not been clear.
The construction of the higher spin fields is required for the full
realization of the  quantum field theories
on quantum spheres.

In this letter, we deal with the quantum $S^2$ and construct Dirac
operators. The advantage of treating the quantum two sphere is that
this manifold is not only the simplest compact quantum manifold in even
dimensions where we can discuss the chiral fermions, but we can use
well known properties of representation theories of quantum $SU(2)_q$.

The algebra $SU(2)_q$\cite{Dri}\cite{Jim} is generated by elements $X^\pm$ and
$K= q^{H_0}$
satisfying the following relations
\bea
K X^\pm K^{-1} = q^\pm X^\pm , \qquad
[X^+, X^- ] = { K - K^{-1} \over q^\oot - q^{- \oot} }.
\ena
A structure of a Hopf algebra is given by
\bea
\Delta (X^\pm) = X^\pm \otimes K^{-\oot} + K^\oot \otimes X^\pm,\qquad
\Delta
(K) = K \otimes K.
\ena
Our convention of this structure describe the one of $SU(2)_{q^{-1}}$
in the convention of ref.\cite{KR}.

The generators of $SU(2)_q$ act on the spin $j$ representation in the
following way,
\bea
X^\pm \jm jm &=& ( \qint {j\mp m} \qint {j\pm m+1})^\oot \jm j{m\pm1}\\
K \jm jm &=& q^m \jm jm,
\ena
where $-j \leq m \leq j, 2m \equiv 2j($mod$2)$ and,
\bea
 \qint n \equiv { q^{n \over 2} - q^{-{n \over2}} \over q^\oot -
q^{-\oot}}.
\ena

We introduce  coordinates of quantum sphere\cite{Pod} as an spin $1$
representation of $SU(2)_q$ which will be denoted by $e_m, m=-1,0,1$.
The commutation relations among coordinates can be defined by requiring
that total spin $1$ states cannot be constructed out of the product of
the
coordinates, which implies
\bea
\sum_{m'} \threej 111{m'}{m-m'}m e_{m'}e_{m-m'} = 0,
\ena
where $\threej {j_1}{j_2}{j_3}{m_1}{m_2}{m_3}$ represents the $q$
analog
of Clebsh-Gordan Coefficient\cite{KR}.  For later convenience, we list the
following
explicit form:
\bea
\threej 1jj0mm &=& \lbrace q^{(j-m+1)/2} \qint {j+m} - q^{-(j+m+1)/2}
\qint
{j-m} \rbrace / \sqrt{ \qint {2j} \qint {2j+2}}, \\
\threej 1jj1{m-1}m &=& - q^{-m/2} \left( { \qint 2 \qint {j+m}
\qint{j-m+1}
\over
\qint{2j} \qint{2j+2}} \right)^{1/2}, \\
\threej 1jj{-1}{m+1}m &=& q^{-m/2} \left( { \qint 2 \qint{j-m}
\qint{j+m+1}
\over
\qint {2j} \qint {2j+2}}\right)^{1/2}. \\
\ena
The equation of the sphere can be expressed as
\bea
e \cdot e = r^2,
\ena
where $a \cdot b$ for spin $j$ representations $a_m, b_m$ is defined by
\bea
a \cdot b &\equiv& - \sum_m (-1)^{j-m} q^{m/2} a_{-m} b_m \nonumber\\
          &=& - \qint {2j+1}^{1/2} \sum_m \threej jj0{-m}m0 a_{-m}b_m,
\ena

The highest weight state of integer spin $j$ can be written as
$\jm jj = a_j e_1^{j}$, where
the coefficient $a_j$ ,
which we do not need to specify in this letter,
can be determined from the haar measure\cite{Wor}\cite{NM}\cite{Suzuki}.
The states $\jm jm$ can be obtained by using a
relation $(3)$. For example, the state $\jm j{j-1}$ is given by $a_j
q^{-(j-1)/2}
({ \qint {2j} \over \qint 2})^{1/2} e_1^{j-1}e_0 $.

We next introduce the derivatives with respect to the coordinates
satisfying
\bea
\partial_{m_1} e_{m_2} = q^{-1} \sum (R^{1,1})_{m_1,m_2}^{m_1',m_2'}
e_{m_2'}
\partial_{m_1'} + (-1)^{m_2} q^{m_2/2} \delta_{m_1 +m_2,0},
\ena
where $(R^{1,1})_{m_1,m_2}^{m_1',m_2'}$ is  so-called the universal
R-matrix\cite{KR} between the spin $1$ states. The element can be written
in terms of the  $3j$- symbol as follows:
\bea
( R^{j_1,j_2} )_{m_1,m_2}^{m_1',m_2'} & = & \sum_j (-1)^{j_1+j_2-j}
q^{-\lbrace j_1(j_1 +1) + j_2 (j_2 +1) - j(j+1) \rbrace /2 } \nonumber
\\
& & \times {\threej {j_1}{j_2}{j}{m_1}{m_2}{m}}
           {\threej {j_2}{j_1}{j}{m_2'}{m_1'}{m}},
\ena
in which $m_1 + m_2 = m_1' + m_2' = m$.
Using these derivatives, we can define the operator of orbital angular
momentum $J_m$ as
\bea
J_m = \left({\qint 4 \over \qint 2} \right)^{1/2} \sum_{m_1} \threej
111{m_1}{m-m_1}m e_{m_1} \Lambda
\partial_{m-m_1},
\ena
where $\Lambda$ is the dilatation operator\cite{Ogi}\cite{OZ}\cite{Fio} whose
action to $e_m$ and
$\partial_m$ is given by
\bea
\Lambda e_m = q e_m \Lambda, \qquad \Lambda \partial_m = q^{-1}
\partial_m
\Lambda.
\ena
The introduction of this operator is required to ensure an relation
\bea
[J_m, r^2] =0.
\ena

Since each operator $e_m$ and $\partial_m$ changes the spin of the
state
by $\pm1$, the operator $J_m$ does not change the spin of the states.
(The total change of spin by $\pm2$ is forbidden.)
Therefore, the operators $J_m$ can be called as quantum rotation
operators.
The action of this operator on the
states $\jm jm$ can be specified as
\bea
J_m \jm j{m'} = { \qint {2j}^{1/2} \qint {2j+2}^{1/2} \over \qint
2}
 \threej 1jjm{m'}{m+m'} \jm j{m+m'},
\ena
where we have
obtained the  coefficient by acting $J_0$ to the highest weight state
 because the $m$ dependence should be included only in the
$3j$-symbols.
We can define a second order Casimir operator by using $J_m$ as
\bea
C = J \cdot J,
\ena
where the product is defined in $(12)$.
By using $(18)$ and relations
\bea
\threej {j_1}{j_2}j{m_1}{m_2}m = (-1)^{j_2 - j - m_1}
q^{-m_1/2}\left({\qint {2j+1} \over \qint {2j_2+1}}\right)^{1/2}
\threej {j_1}j{j_2}{-m_1}m{m_2} ,
\ena
\bea
\sum_{m_1,m_2} \threej {j_1}{j_2}j{m_1}{m_2}m \threej
{j_1}{j_2}{j'}{m_1}{m_2}{m'} = \delta_{j,j'}\delta_{m,m'},
\ena
for $\vert j_1 - j_2 \vert \leq j \leq j_1 + j_2$,
we find the value of the Casimir operator as
\bea
C \jm jm = { \qint {2j} \qint {2j+2} \over \qint 2^2} \jm jm .
\ena
Using this Casimir operator,
we can construct an equation of the scalar field of mass $m$
on this Euclidean quantum space as
\bea
\left({C \over r^2} + m^2\right) \phi (e) = 0.
\ena

Our next question is whether chiral fermions can exist on this quantum
two sphere.

To begin with, we operatorially introduce the spin $1/2$ spinors as
\bea
\jm {1/2}{1/2} &=& c_{1/2} \ket 0, \nonumber\\
\jm {1/2}{-1/2} &=& c_{-1/2} \ket 0.
\ena
It is not necessary to introduce commutation rule among the operators
$c$'s since we only consider one particle states with respect to
$c$'s..
We next define the conjugate of the fields $c_m$ preserving the
$SU(2)_q$
invariant:
\bea
b_{m_1} c_{m_2} \ket 0 = \qint 2^{1/2} \threej {\oot}{\oot}0{m_1}{m_2}0
\ket 0.
\ena
Explicitly, we obtain
\bea
b_{-1/2} c_{1/2} \ket 0 &=& q^\oof \ket0, \qquad b_{1/2}c_{-1/2} \ket 0
= -
q^\oof \ket 0, \\
b_{1/2} c_{1/2} \ket 0 &=& b_{-1/2} c_{-1/2} \ket 0 = 0.
\ena
{}From these operators, we can introduce q-analog of Pauli matrices as
\bea
\sigma_m = [2]^{1/2} \sum_{m_1} \threej {\oot}{\oot}1{m_1}{m-m_1}m  c_{m_1}
b_{m-m_1},
\ena
which implies
\bea
\sigma_1 &=& [2]^{1/2}c_{1/2}b_{1/2}, \nonumber\\
\sigma_0 &=&  q^{-1/4}c_{-1/2}b_{1/2} +
q^{1/4}c_{1/2}b_{-1/2} , \nonumber\\
\sigma_{-1} &=& [2]^{1/2} c_{-1/2}b_{-1/2}.
\ena
We also define a singlet operator as
\bea
O &=& c \cdot b \nonumber\\
  &=& -q^{1/4} c_{-1/2}b_{1/2} + q^{-1/4}c_{1/2}b_{-1/2}.
\ena
The action of this operator on $c_\pm \ket 0$ is just an identity
operator.
The $q$-deformed Pauli-matrices form an algebra:
\bea
\sigma_{m_1} \sigma_{m_2} = { \qint 4^{1/2} \over \qint 2^{1/2}} \threej
111{m_1}{m_2}{m_1 + m_2} \sigma_{m_1+m_2}
+ (-1)^{m_1} q^{-m_1/2} \delta_{m_1+m_2,0}O,
\ena
which is the q-analog of the well known relation
$\sigma^i \sigma^j = i \varepsilon_{ijk}\sigma^k + \delta^{ij}.$

Since the general states can be expressed as products of the
coordinates and
$c_m$,
we need to specify the commutation relations between $e_m$ and
$c_{m'}$.
A consistent choice is to use the universal R-matrix\cite{KR} with respect to
spin $1/2$
and spin $1$:
\bea
c_{m_1}e_{m_2} = (R^{1/2,1})_{m_1,m_2}^{m_1',m_2'} e_{m_2'} c_{m_1'}.
\ena
We can find that $c_m$ commutes with the radius $r^2$.
As for the commutation relation between $b_{m_1}$ and $e_{m_2}$, we
prepare
two type of conjugate of $c_m$.  One is defined to commute with
$e_m$ by
$R^{-1}$ matrix which we call $b_m$, the other is to commute
with
$e_m$ by R-matrix, which will be  $b'_m$.
Correspondingly,
the Pauli-matrices and operator $(30)$ constructed by $c_m$ and $b_m$
will
be denoted by $\sigma_m$ and $O$ and the operators constructed by $c$
and
$b_m'$ will be called $\sigma_m'$ and $O'$.

It seems instructive to write down the commutation relations explicitly:
\bea
\cp \ep &=& \qh \ep \cp, \qquad \cm \emi = \qh \emi \cm, \nonumber\\
\cp \ez &=& \ez \cp + \qint 2^{1/2} \qmq \qhmh \ep \cm, \nonumber\\
\cm \ep &=& \qh \ep \cm, \qquad \cm \ez = \ez \cm, \nonumber\\
\cp \emi &=& \qmh \emi \cp + \qint 2^{1/2} \qmq \qhmh \ez \cm,\\
\bp \ep &=& \qmh \ep \bp, \qquad \bm \emi = \qmh \emi \cm, \nonumber\\
\bm \ez &=& \ez \bm - \qint 2^{1/2} \qq \qhmh \emi \bp, \nonumber\\
\bp \emi &=& \qh \emi \bp, \qquad \bp \ez = \ez \bp, \nonumber\\
\bm \ep &=& \qh \ep \bm - \qint {2}^{1/2} \qq \qhmh \ez \bp,
\ena
The states constructed by the product of spin $j$ representation of
the coordinate space and spin $1/2$ representation split into the
total spin $j+{1/2}$ and spin $j-1/2$.
The highest weight states can be written as
\bea
\jjm j{\jph}{\jph} &=& \jm jj c_{1/2} \ket 0, \\
\jjm j{\jmh}{\jmh} &=& \nonumber\\
{1 \over \qint {2j+1}^{1/2} } &\lbrace& q^{j/2} \jm j{j-1} c_{1/2,1/2}
-
q^{-1/4} \qint {2j}^{1/2} \jm jj c_{-1/2} \rbrace  \ket 0 .
\ena
The other descendants can be derived by using a relation $(3)$.

We are going to consider the action of $\sigma_m$ on these states.
Since
$\sigma_m$ does not change bosonic spin and preserve $SU(2)_q$
invariance,
we have
\bea
\sigma_m \jjm j{\jph}{m'} = &b_{++}& \threej 1\jph\jph{m}{m'}{m+m'}
\jjm
j\jph{m+m'} \nonumber\\
                            &+& b_{+-} \threej 1\jph\jmh{m}{m'}{m+m'}
\jjm
j\jmh{m+m'}.
\ena
The coefficients $b_{++}$ and $b_{+1}$ can be determined by considering
$\sigma_0 \jjm j\jph\jph$ and $\sigma_{-1} \jjm j\jph\jph $, and we get
\bea
b_{++} &=& \left({ \qint {2j+3} \over \qint {2j+1}
}\right)^{1/2}q^j, \\
b_{+-} &=& -\left( { \qint {2j+2} [2] \over \qint {2j+1} } \right)^{1/2}
q^{-1/2}.
\ena
Quite similarly, we have
\bea
\sigma_m \jjm j\jmh{m'} = &b_{-+}& \threej 1\jmh\jph{m}{m'}{m+m'} \jjm
j\jph{m+m'} \nonumber\\
                        &+& b_{--} \threej 1\jmh\jmh{m}{m'}{m+m'} \jjm
j\jmh{m+m'},
\ena
where
\bea
b_{-+} &=& \left({ \qint {2j} [2] \over \qint {2j+1}}\right)^{1/2}
q^{-1/2},
\\
b_{--} &=& - \left({\qint {2j-1} \over \qint {2j+1} }\right)^{1/2}
q^{-j-1}.
\ena
As for the action of $J_m$, we get the following relations:
\bea
J_m \jjm j{\jph}{m'} = &a_{++}& \threej 1\jph\jph{m}{m'}{m+m'} \jjm
j\jph{m+m'}
\nonumber\\
                            &+& a_{+-} \threej 1\jph\jmh{m}{m'}{m+m'}
\jjm
j\jmh{m+m'}.\\
J_m \jjm j\jmh{m'} = &a_{-+}& \threej 1\jmh\jph{m}{m'}{m+m'} \jjm
j\jph{m+m'}
\nonumber\\
                        &+& a_{--} \threej 1\jmh\jmh{m}{m'}{m+m'} \jjm
j\jmh{m+m'},
\ena
where the coefficients can be obtained as
\bea
a_{++} &=& { \qint {2j} \qint {2j+3}^{1/2} \over \qint 2 \qint
{2j+1}^{1/2}}, \qquad a_{+-} = { \qint {2j+2}^{1/2} \over \qinth
{2j+1} \qinth {2}}, \\
a_{-+} &=& - { \qinth {2j} \over \qinth {2j+1} \qinth {2}}, \qquad a_{--} = {
\qint {2j+2}
\qinth {2j-1} \over \qint 2 \qinth {2j+1}}.
\ena
{}From these relations, we can define the operator representing total
angular
momentum as
an operator which does not change the total spin. Indeed by  defining
the
operator $J_m^{tot}$ as
\bea
J^{tot}_m = q^{-1/2} J_m + {1 \over [2]} \sigma_m,
\ena
and using $(37)(40)(43)$ and $(44)$, we obtain
\bea
J^{tot}_m \jjm j\jph{m'} &=& \nonumber\\
\left( { \qint {2j+1} \qint {2j+3} \over \qint 2^2}\right)^{1/2} &\threej
1\jph\jph{m}{m'}{m+m'}& \jjm j\jph{m+m'},\\
J^{tot}_m \jjm j\jmh{m'} &=& \nonumber\\
\left({ \qint {2j-1} \qint {2j+1} \over \qint 2^2}\right)^{1/2} & \threej
1\jmh\jmh{m}{m'}{m+m'}& \jjm j\jmh{m+m'}.
\ena

 From these relations,
we can find the value of the second order Casimir operator $C^{tot} = J^{tot}
\cdot J^{tot}$
as
\bea
C^{tot} \jjm j\jph{m} &=& { \qint {2j+1} \qint {2j+3} \over \qint 2^2}
\jjm
j\jph{m},\\
C^{tot} \jjm j\jmh{m} &=& { \qint {2j-1} \qint {2j+1} \over \qint 2^2}
\jjm
j\jmh{m},
\ena
Note also that the operator $O$ defined in $(30)$ is no longer an
identity
operator.
We obtain
\bea
O \jjm j\jph{m} &=& q^j \jjm j\jph{m},\\
O \jjm j\jmh{m} &=& q^{-j-1} \jjm j\jmh{m},
\ena
whereas $O' \jjm j{j\pm 1/2}m = \jjm j{j\pm 1/2}m.$
We can also find that the action of the operator $\sigma_m'$ on states
can be written as $\sigma_m' = O^{-1} \sigma_m$.

Another  Casimir operator can be constructed from the product of
$J_m$ and $\sigma_m$ as $J \cdot \sigma$,
which is the first-order operator with respect to the derivatives of the
coordinates.
The value of this Casimir operator can be
obtained using relations $(37),(40),(43)$ and $(44)$:
\def\jjph{\jjm j\jph{m}}
\def\jjmh{\jjm j\jmh{m}}

\bea
J \cdot \sigma \jjph &=& \big( a_{++}b_{++} - b_{+-}a_{-+}{ \qinth {2j}
\over
\qinth {2j+2}}\big) \jjph, \nonumber\\
                     &=&  { \qint {2j} \over \qint 2} q^{j+1},\\
J \cdot \sigma \jjmh &=& \big( a_{--}b_{--} - b_{-+}a_{+-}{ \qinth
{2j+2} \over
\qinth {2j}}\big) \jjmh \nonumber\\
                     &=& - { \qint {2j+2} \over \qint 2} q^{-j}\jjmh .
\ena
Before trying to construct a Dirac operator, we first consider
$\gamma_5(\gamma_3)$.
In two dimensions, we usually take $\sigma^i$ $(i = 1,2)$ as gamma
matrices for
the coordinates
and $\sigma_3$ can be used as an $\gamma_5$ element. In our
construction,
we have already used all the Pauli matrices. This fact implies
that one of the matrices we have used are redundant.
As our $\gamma_5$, we are going to choose the $\gamma$ matrix
corresponding to
the direction of the radius. In other words,
we take
\bea
\gamma_5 = i q^{-1} {e \cdot \sigma' \over \sqrt{r^2}},
\ena
where $\sigma_m$ represents the Pauli matrix constructed from $b'_m$
and
$c_m$,
and the normalization is chosen to ensure $ \gamma_5^2 = -1$.
This operator change the orbital angular momentum by $\pm 1$ but does
not change the total spin of the states.
We therefore find
\bea
\gamma_5 \jjph &=& \alpha_j \jjm {j+1}{j+1/2}m , \\
\gamma_5 \jjmh &=& \beta_j \jjm {j-1}{j+1/2}m ,
\ena
where $\alpha_j$ and $\beta_j$ are some coefficients which we do not
need to
specify for the foregoing discussion.
Note also that this operator commutes with the total second order
Casimir operator $C_{tot}$.
We are now going to construct Dirac operator. Our ansatz for the
massless dirac operator is of the form
\bea
r \gamma \cdot \nabla = O^{-1}( J \cdot \sigma + a),
\ena
where $r$ is the radius of the sphere.
We will fix the value of $a$ be requiring that this operator
anticommute with
$\gamma_5$ for states $\jjph$ and $\jjmh$. It is not a priori clear
that
we can find the same $a$ for these two states but we have $a =1$.
The value of the Dirac operator on the states is obtained as
\bea
r\gamma \cdot \nabla \jjph &=& { \qint {2j+2} \over \qint 2} \jjph,\\
r\gamma \cdot \nabla \jjmh &=& -{ \qint {2j} \over \qint 2} \jjmh.
\ena
{}From these we can find the relation to the total second order
Casimir operator
as
\bea
C^{tot}  + { 1 \over \qint 2^2} = ( r \gamma \cdot \nabla)^2.
\ena
Note also that the massive Dirac operator $ \gamma \cdot \nabla + im$
no longer
anticommute with $\gamma_5$.

We have constructed Dirac operator on the quantum two spheres.  We
expect
that the
fields with higher spins can be described using our approach.
For physical application, the quantum four sphere is of particular
interest. Indeed, there is no difficulty to extend the result of this
paper to the quantum four spheres. We will discuss this issue in a
separate  paper\cite{OS}.


\newpage
\begin{thebibliography}{99}


\bibitem{Dri}
V. G. Drinfeld,
Sov. Math. Dokl. {\bf 32} (1985) 1254.

\bibitem{Jim}
M. Jimbo,
Lett. Math. Phys. {\bf 10} (1985) 63.

\bibitem{Wor}
S. L. Woronowicz,
Publ. RIMS, Kyoto Univ. {\bf 23} (1987) 117.


\bibitem{Man}
Yu. I. Manin,
$Quantum\ Groups\ and\ Non\!-\!Commutative\ Geometry$,
Preprint Montreal University, CRM-1561 (1988)

\bibitem{Pod}
P. Podle\`s,
Lett. Math. Phys. {\bf 14} (1987) 193

\bibitem{WZ}
J. Wess, B.Zumino,
Nuclear Physics B (Proc. Suppl.) {\bf 18B} (1990) 302.

\bibitem{CSSW}
U. Carow-Watamura, M. Schlieker, M. Scholl, S. Watamura,
Z. Phys. C-Particles and Fields {\bf 48} (1990) 159;
U. Carow-Watamura, M. Schlieker, S. Watamura,
\CMP{142} (1991) 605.

\bibitem{Suzuki}
H.Suzuki,
Prog. of Theor. Phys.91(1994)379.

\bibitem{RTF}
N. Yu. Reshetikhin, L. A. Takhtadzhyan, and L. D. Faddeev,
Leningrad Math. J. {\bf 1} (1990) 193.


\bibitem{KR}
A. N. Kirillov, N. Yu. Reshetikhin,
$Representations\ of\ the\ Algebra\ U_{q}(sl(2)),\ q-Orthogonal\
Polynomials\
and\ Invariants\ of\ Links$,
preprint 1988.

\bibitem{NM}
M.Noumi and K.Mimachi,
Commun.Math.Phys.128(1990)521.


\bibitem{Ogi}
O.Ogievetsky,
Lett.Math.Phys.24(1992),245.

\bibitem{OZ}
O.Ogievetsky and B.Zumino,
Lett.Math.Phys.25(1992),121.

\bibitem{Fio}
G. Fiore,
preprint SISSA-90-93-EP, hep-th/9403033

\bibitem{OS}
K.Ohta and H.Suzuki, in preparation.


\end{thebibliography}
\end{document}